\documentclass[11pt,a4paper]{article}
\usepackage[T1]{fontenc}
\usepackage[utf8]{inputenc}
\usepackage{authblk}

\usepackage{geometry}
\usepackage{amsfonts,amsmath,amssymb,amsthm,nicefrac,amscd}
\usepackage{ulem}
\usepackage{cancel}
\usepackage{yhmath}
\usepackage{arcs}
\usepackage{graphicx}
\usepackage{url}
\usepackage{physics}
\usepackage{mathtools} 

\usepackage{cite} 
\usepackage{times} 

\usepackage{xcolor} 

\definecolor{dgreen}{rgb}{0.1, 0.4, 0.2}

\usepackage{caption}
\usepackage{subcaption}

\usepackage[bookmarks=false]{hyperref}
\hypersetup{%
    colorlinks=true,        
    linkcolor=blue,          
    citecolor=blue,         
    urlcolor=blue           
    }
\allowdisplaybreaks 

\providecommand{\keywords}[1]
{
  \small	
  \textbf{\textit{Keywords: \quad 
  }} #1
}

\title{\color{blue}\bf X-states of a qubit pair of double classicality}
\author[1,2,3]{Arsen Khvedelidze} 
\author[4,5]{Dimitar Mladenov}
\author[3,6]{Astghik Torosyan}

\affil[1]{A.~Razmadze Mathematical Institute,
Iv.~Javakhishvili Tbilisi State University, Tbilisi, Georgia}
\affil[2]{Institute of Quantum Physics and Engineering Technologies, Georgian Technical University, Tbilisi, Georgia}
\affil[3]{Laboratory of Information Technologies, Joint Institute for Nuclear Research, Dubna, Russia} 
\affil[4]{Faculty of Physics, Sofia University "St. Kliment Ohridski", 5 James Bourchier Blvd, 1164 Sofia, Bulgaria} 
\affil[5]{Institute of Nuclear Research and Nuclear Energy, Bulgarian Academy of Sciences, 72 Tzarigradsko Shaussee, 1784 Sofia, Bulgaria} 
\affil[6]{A.I.~Alikhanyan National Science Laboratory (YerPhI), Yerevan, Armenia}

\date{ } 

\begin{document}

\maketitle

\begin{abstract}
A special class of states of 2-qubits  which are simultaneously separable and have  positive semidefinite Wigner functions is described.
\end{abstract} 


\newpage
\tableofcontents

\newpage 


\section{Introduction}

\paragraph{Two aspects of nonclassicality in quantum systems.} A crucial goal of quantum technologies is to utilise deviations of the quantum system from their classical counterparts as a resource that allows for significant improvements in the effectiveness of classical devices. The \textit{entanglement and negativity of the quasi-probability distributions of quantum states} are among the commonly accepted resources for achieving a quantum advantage.
Bearing in mind this goal, we introduce a special class of resourceful two-qubit states as a complement to the intersection of the convex sets of separable states and states whose Wigner functions are nonnegative. The latter play a role of the ``free'' states of quantum resource theory (cf. \cite{Deneris2025} and references therein).  
In our consideration, we follow the phase-space formulation of finite-dimensional quantum systems in a spirit of the Stratonovich-Weyl (SW) correspondence \cite{Stratonovich1957,BrifMann1999,KhA2018}.  More specifically, since entanglement is a phenomenon occurring in composite systems, we use a generalised SW method to construct the Wigner function that takes into account a composite nature of quantum states through imposed algebraic conditions on the spectrum of the corresponding SW kernels (see details in \cite{AKH2023,AKHR2025}).

\section{Classifying states of N\--dimensional quantum system}
\label{eq:SepvsPos}

Here,   starting with the definitions of two convex subspaces of the state space $\mathfrak{P}_N$ of an N--dimensional quantum system: \textit{ the set of mixed separable states $\mathfrak{S}_N$ and the set of states $\mathfrak{C}^{(+)}_N$} whose Wigner functions are nonnegative,
we introduce the set of \textit{double classicality states} $\mathfrak{C^{(++)}}_N \subseteq\mathfrak{S}_N\cap\mathfrak{P}^{(+)}_N\,.$

\paragraph{Classifying states into separable vs. entangled sets.}
Let $\mathfrak{P}_N$ be the
state space of an $N$-dimensional bipartite system composed of two subsystems, $A$ and $B$. The associated Hilbert space of the whole
system is prescribed by the tensor product of the subsystems' Hilbert spaces, $\mathcal{H}_{A\times B} \subseteq \mathcal{H}_{A}\otimes \mathcal{H}_B\,. $ Attributing
the tensorial structure to the Hilbert space $\mathcal{H}_{{A}\times{B}}$
sets apart from the global properties
of total system
(like its dimension, 
$N =N_AN_B$\,, where  $\dim{\mathcal{H}_{A}}=N_A\,,$  $\dim{\mathcal{H}}_{B}=N_B$ 
 )
also  provides possibility to divide  the state space $\mathfrak{P}_N$ into two complementary sets, 
the family of \textit{separable mixed states $\mathfrak{S}_N \subset \mathfrak{P}_N$} and its complement \---
the set of \textit{entangled mixed states}. The set $\mathfrak{S}_N$ is defined by a convex combination of tensor products of the states of subsystems
$\varrho_A^k$ and $\varrho_B^k$:
\begin{eqnarray}
 \mathfrak{S}_N:\quad  \{\,   
 \varrho^k_A \in\mathfrak{P}_{N_A},\,
 \varrho^k_B \in\mathfrak{P}_{N_B}
 \,|\,   
\mbox{conv}(\varrho_A^k\otimes \varrho^k_B)\, \}\,.
\end{eqnarray}

\paragraph{Classifying states according to  the  sign of Wigner functions.} Let $W_\varrho (\boldsymbol z)$
be the Wigner function of a state $\varrho$ of the composite system $\mathcal{H}_{A\times B}.$
\footnote{For readers convenience, we briefly recall some necessary notions from 
\cite{AKH2023}.   $W_\varrho (\boldsymbol z)$ is a dual pairing
between a density matrix $\varrho$ and the Stratonovich-Weyl kernel $\Delta(\boldsymbol{z})\,$:
\begin{equation}
\label{eq:WignerFunction}
W_\varrho(\boldsymbol{z})= \mbox{tr}\left[\varrho\, \Delta(\boldsymbol{z})\right]\,, \qquad  
\boldsymbol{z}=(z_1, z_2, \dots, z_d )\, 
\in \Omega_N\,. 
\end{equation} 
If the N\--level system is treated as an elementary one, then $\Delta(\Omega_N)\in \mathfrak{P}^\ast_N\,,$ where $ \mathfrak{P}^\ast_N\,$ is the space of Hermitian $N\times{N}$ matrices with the spectrum $\mbox{\bf spec}(\Delta_N)=(\pi_1,\pi_2, \dots, \pi_N) $ specified by the equations:
\begin{equation}
\label{eq:SWMaster}
\mathfrak{P}^\ast_N :\quad 
\sum_{i=1}^N\,\pi_i=1\,, \quad
 \sum_{i=1}^N\,\pi_i^2=N\,.
\end{equation}
Both the phase space \(\Omega_N \) and the set of solutions to (\ref{eq:SWMaster}), i.e. the moduli, are determined by the isotropy group of SW kernel as:
\(
\Omega_N={SU(N)}/{\mathrm{Iso_{{}_{SU(N)}}}(\Delta)}\,,  
\ 
\mathcal{P}_N={\mathfrak{P}^\ast_N}/{\mathrm{Iso_{{}_{SU(N)}}}(\Delta)}\,.\) If the $N\--$level system is known to be composite, then the SW kernel must satisfy additional constraints on the partially reduced SW kernels 
$\Delta_A=\mbox{tr}_B\Delta$ and 
$\Delta_B=\mbox{tr}_A \Delta$, 
\begin{equation}
\label{eq:SWMasterCom}
\mathfrak{P}^\ast_{A\times{B}}:
\quad
\mbox{tr}_{{}_A}(
\Delta_A)^2=N_{{}_A}\,, \quad  
\mbox{tr}_{{}_B}(\Delta_B)^2=N_{{}_B}
\,.
\end{equation}
Similarly, the phase space $\Omega_{A\times{B}}$ and the moduli space  $\mathcal{P}_{A\times{B}}$
are modified. They are now determined by the Local Unitary (LU)  subgroup:
\(
 \mathrm{LU}=SU(N_A)\times SU(N_B) \subset{SU(N)}\,  
\) 
as the cosets 
\( \Omega_{A\times{B}}={{\mathrm{LU}}}/{\mathrm{Iso_{{}_{LU}}}(\Delta)} \,,
 \) and \(
 \mathcal{P}_{A\times{B}}=
{\mathfrak{P}^\ast_{A\times{B}}}/{\mathrm{LU}}
\) respectively. 
} 
Selecting the states whose WF is non-negative, we define the subset 
$\mathfrak{C}_N^{(+)}\subseteq\mathfrak{P}_N$:
\begin{eqnarray}
\label{eq:ClassicalS}
\mathfrak{C}_N^{(+)} &=& \{\, \varrho \in \mathfrak{P}_N, \, \Delta \in \mathfrak{P}^\ast_{A\times B}\ | \  W_\varrho(z) \geq 0\,, \quad  \forall z\in \Omega_N \, \}\,.    
\end{eqnarray}
We call the elements of $\mathfrak{C}_N^{(+)}$ 
``classical states'' and emphasise that the associated WFs are proper statistical probability distributions. 

\paragraph{Double classicality states.} The intersection of two convex bodies, the separable $\mathfrak{S}_N$ and WF positive semidefinite ones, defines the set of states
\begin{equation}
\mathfrak{C}^{(++)}_{N}=\mathfrak{S}_N\,\cap\,\mathfrak{C}^{(+)}_N\,, 
\end{equation}
which we call the states of double classicality.

\section{Separable and absolute separable X-states of 2-qubits}
\label{sec:SepX}

A density matrix of a pair of qubits is called an $X$\--state if it belongs to the subset $\mathfrak{P}_X \subset \mathfrak{P}_4$  of matrices whose shape resembles the Latin letter ``X'':
\begin{equation}
\label{eq:XM2}
\varrho_{X}:=
\left(
\begin{array}{cccc}
\varrho_{11}& 0 &0& \varrho_{14}\\
0&\varrho_{22} &\varrho_{23}& 0\\
0&\overline{\varrho}_{23} &\varrho_{33}& 0\\
\overline{\varrho}_{14}& 0 &0& \varrho_{44}
\end{array}
\right)
\,.
\end{equation}
The state $\varrho_X$ is similar to a block-diagonal matrix and is therefore unitary equivalent to a diagonal matrix
\(
\label{eq:SVDeltaX}
\varrho_X= U
\mbox{diag}(
\boldsymbol{r}^\uparrow
)
U^\dagger\,, 
\) with the following unitary factor
\begin{equation}
\label{eq:UForm}
    U=P\left(
\begin{array}{c|c}
U_1& 0 \\
\hline
0&U_2 
\end{array}
\right)Q\,, 
\end{equation}
where $P$ and $Q$ stand for the permutation matrices that perform the transposition of rows and columns, 
$U_1, U_2 \in SU(2)/U(1)\,.$ The latter can be parameterized by the Eulerian angles $\phi_1,\phi_2 \in [0, \pi]\,, \psi_1,\psi_2 \in [0, 2\pi]$: 
\begin{equation}\label{eq:UV}
U_1 =e^{i \displaystyle{\frac{\psi_1}{2}\sigma_3}}e^{i \displaystyle{\frac{\phi_1}{2}\sigma_2}}\,,
\qquad
U_2=e^{i \displaystyle{\frac{\psi_2}{2}\sigma_3}}e^{i \displaystyle{\frac{\phi_2}{2}\sigma_2}}\,.
\end{equation}

\paragraph{Separable states $\mathfrak{S}_X$.}
Applying the Peres-Horodecki criterion \cite{BZ} to the X\--states presented in the decomposition described above results in the following conditions on the density matrix spectrum and
two Euler angles:
\begin{eqnarray}
\label{eq:1posTREig}
&&(r_1-r_2)^2\cos^2\phi_1+(r_3-r_4)^2 \sin^2\phi_2 \leq (r_1+r_2)^2,
\\
&&
(r_3-r_4)^2\cos^2\phi_2+(r_1-r_2)^2 \sin^2\phi_1 \leq (r_3+r_4)^2.
\label{eq:2posTREig}
\end{eqnarray}

\begin{figure}
 \centering
 \begin{subfigure}[b]{0.46\linewidth}
\includegraphics[scale=0.5
]{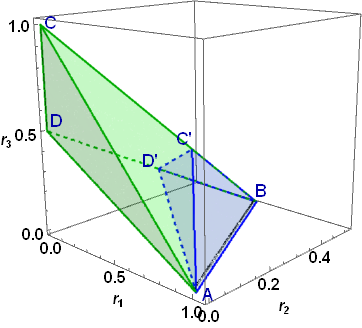}
  \end{subfigure}
  \hspace{0.8cm}
  \begin{subfigure}[b]{0.46\linewidth}
\includegraphics[scale=0.5
]{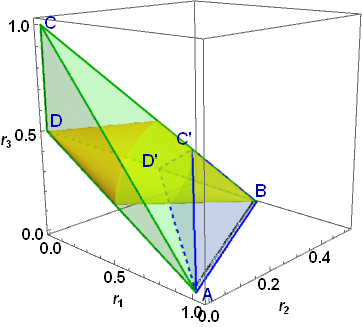}
  \end{subfigure}
\caption{Left: Tetrahedron  $ABCD$\--- the simplex of partially ordered eigenvalues satisfying $1>r_1>r_2>0,\, 1>r_3>r_4>0$, while  $ABC'D'$ \--- the fundamental simplex with 
$1 \geq r_1 \geq r_2 \geq r_3 \geq r_4 \geq  0\,$.
Right: Intersection of absolutely separable states with the fundamental simplex. 
}{ \label{Fig:PartOrderedSimplex}}
\end{figure}

\paragraph{Absolutely separable states.}There exists a  special family  of   ``absolutely  separable'' $X$-states that are separable for all angles $\phi_1$ and $\phi_2$:
\begin{eqnarray}\label{eq:absSepX}
(r_1-r_2)^2  \leq  4r_3r_4\,,\quad 
(r_3-r_4) ^2  \leq 4r_1r_2\,.
\end{eqnarray}
The absolutely separable states geometrically represent the convex body obtained by the union of two cones whose apexes coincide with two vertices of the 3-simplex and their bases are glued together; see Figure \ref{Fig:PartOrderedSimplex}.   

\section{WF positivity for quatrit and 2-qubits in X-states}
\label{sec:WFposX}

\paragraph{WF positivity polytope.}
The Wigner function of a mixed N-level state $\varrho$ is bounded, its bounds are determined by the spectrum
$\mbox{spec}(\varrho)=\{\boldsymbol{r}^\uparrow\} \,$ and the SW kernel spectrum $\mbox{spec}(\Delta)=\{\boldsymbol{\pi}^\uparrow\}$:
\begin{equation}
\label{eq:Bounds}
 (\boldsymbol{r}^\uparrow\cdot
\boldsymbol{\pi}^\downarrow) \leq   W_\varrho(\Omega_4)\leq (\boldsymbol{r}^\uparrow\cdot
\boldsymbol{\pi}^\uparrow)\,.
\end{equation}
In (\ref{eq:Bounds}) the superscripts $\uparrow(\downarrow)$ over the $N$-tuple $\boldsymbol{x}=\{x_1,x_2, \dots, x_N\}$ denote the descending (ascending) orderings
of its elements.
Hence, according to (\ref{eq:Bounds}), the subset $\mathfrak{C}^{(+)}$ consists of density matrices whose eigenvalues  lie within 
the \textit{WF positivity polytope}, that is, the intersection of the $(N-1)$-simplex of eigenvalues with the supporting hyperplane 
\cite{KhT2023}: 
\begin{equation}
\label{eq:WFPosHyper}
    (\boldsymbol{r}^\uparrow\cdot
\boldsymbol{\pi}^\uparrow) \geq 0\,.
\end{equation}

\paragraph{SW kernels of X\--states.}
The WF of an X\-state for both elementary or composite systems has an SW kernel of the form: 
\begin{equation}
\label{eq:SWK2}
\Delta_{X}:=
\left(
\begin{array}{cccc}
\Delta_{11}& 0 &0& \Delta_{14}\\
0&\Delta_{22} &\Delta_{23}& 0\\
0&\overline{\Delta}_{23} &\Delta_{33}& 0\\
\overline{\Delta}_{14}& 0 &0& \Delta_{44}
\end{array}
\right)
\,.
\end{equation}
However, the  moduli space depends on the compositeness of the system. Below, an explicit representation of the moduli spaces for 4-level system (quatrit) and 2-qubits will be given and the WF positivity polytope will be described. 

\paragraph{The moduli space of quatrit.}The moduli space of a quatrit is 2-parametric. Fixing $\pi_1$ and $\pi_2$ as free moduli, the remaining two eigenvalues can be determined from the master equation (\ref{eq:SWMaster}):
\begin{eqnarray}
\label{eq:specEl1}
\pi_{3,4}&=&\frac{1-\pi_1-
\pi_2 }{2} \pm \frac{1}{2}\sqrt{\,\mathrm{Disc}}\,,
\end{eqnarray}
where $\mathrm{Disc}=7+2(\pi_1+\pi_2-\pi_1\pi_2)-3(\pi_1^2+\pi_2^2)$ and  
the moduli space of the WF of a quatrit represents the domain of the discriminant semi-positivity:
\begin{eqnarray}
\label{eq:moduli4level}
\mathcal{P}_4=\{\pi_1,\pi_2 \in \mathbb{R}^2\,|\, \mathrm{Disc}\geq 0\,\}\,.
\end{eqnarray}

\paragraph{SW kernel of 2 qubits.}
According to  \cite{AKH2023}, if 
the 4-level system is composed from 2-qubits, then the moduli space
(\ref{eq:moduli4level}) is further constrained. 
According to the master equations (\ref{eq:SWMasterCom}), the SW kernel (\ref{eq:SWK2}) obeys the constraints:
\begin{eqnarray}
\label{eq:EQDelta1}
&& \Delta_1+\Delta_2+\Delta_3 +\Delta_4=1\,,\quad
\Delta_1^2+\Delta_2^2+\Delta_3^2 +\Delta_4^2=4-2\,\delta^2\,,\\
&& (\Delta_1+\Delta_2)^2+(\Delta_3+\Delta_4)^2=2, \quad 
(\Delta_1+\Delta_3)^2+(\Delta_2+\Delta_4)^2=2\,,
\label{eq:EQDelta4}
\end{eqnarray}
where $\delta=\sqrt{|\Delta_{14}|^2+|\Delta_{23}|^2}\,.$
Equations (\ref{eq:EQDelta1})-(\ref{eq:EQDelta4}) define a 2-parameter family of SW kernels with the following eigenvalues:
\begin{eqnarray}
\label{eq:specCOM1}
\pi_{1,3} &= &  \frac{1}{4} \pm |\Delta_{14}|+ \frac{1}{4}\sqrt{9-8\,\delta^2 }\,, \\
\pi_{2,4} &= &  \frac{1}{4}\left(1 \pm 2\sqrt{3+4|\Delta_{23}|^2}- \sqrt{9-8\,\delta^2 }\right)\,.
\label{eq:specCOM4}
\end{eqnarray} 
In (\ref{eq:specCOM1}) and (\ref{eq:specCOM4}) the absolute values of the non-diagonal entries of (\ref{eq:SWK2}) represent the  moduli of the SW kernel of 2 qubits 
assuming $\pi_1\geq \pi_2\geq  \pi_3\geq \pi_4: $
\begin{eqnarray}
\mathcal{P}_{2\times 2}: =  
\{\,|\Delta_{14}|<\frac{3}{2\sqrt{2}}\,,  \quad |\Delta_{23}|<\frac{1}{2\sqrt{2}} \sqrt{9-8 |\Delta_{14}|^2}\,\}\,.
\end{eqnarray}

\begin{figure}[ht]
\centering
\begin{subfigure}[b]{0.4\linewidth}
\includegraphics[width=0.9\linewidth]{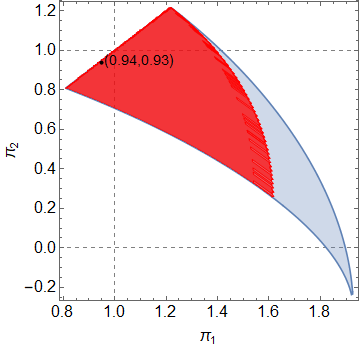}
  \end{subfigure}
  \hspace{0.8cm}
  \begin{subfigure}[b]{0.4\linewidth}
\includegraphics[width=\linewidth]{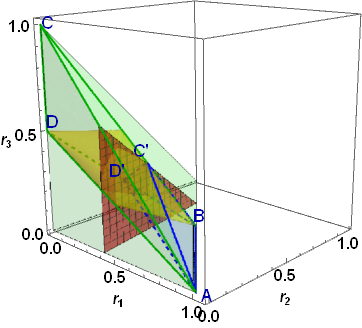}
  \end{subfigure}
\caption{Left: Moduli space of quatrit vs. pair of qubits; 
Right: Typical pattern of intersection of a qubit pair  
Wigner function positivity supporting hyperplane plane (\ref{eq:WFPosHyper}) with the fundamental simplex, ($\pi_1=0.94, \pi_2=0.93, \pi_3=0.51$).}
\label{Fig:WFPolAS}
\end{figure}

Comparing expressions (\ref{eq:specEl1}) and (\ref{eq:specCOM1})-(\ref{eq:specCOM4}), we identify the moduli space of the 2-qubit system as a subset $\mathcal{P}_{2\times 2} \subset \mathcal{P}_{4}\,,$ depicted in Figure \ref{Fig:WFPolAS}.

\section{Concluding remarks}
\label{sec:ComSP}

In the present note, we introduce the notion of doubly classical states as a candidate of ``free'' states in quantum resource theory. The existence of such states follows from a generic geometric structure of the state space;  the sets of separable states $\mathfrak{S}_N$ and classical
states $\mathfrak{C}_N^{(+)}$ are convex subsets of the full state space $\mathfrak{P}_N.$ Therefore, each of these subsets contains an inscribed ball centred in the maximally mixed state
$\varrho_0 =
\frac{1}{N}\mathbb{I}_N\,.$ 
The radius $r_{\mathrm{sep}}$ of the separability ball and the radius
$r_*$ of the Wigner function positivity ball
(the ball of absolute classicality \cite{KhT2023}) are: 
\begin{equation}
    r_{\mathrm{sep}} = \frac{1}{(N-1)}\,,\qquad r_\ast=\frac{\sqrt{N+1}}{N^2-1}\,. 
\end{equation}
Since $r_{\mathrm{sep}} > r_\ast,$  a part of doubly classical states lies entirely within the WF positivity ball. 
Our studies  extend these results; a common locus of two-qubit separable $\mathfrak{S}_4$ and classical  $\mathfrak{C}^{(+)}_4$ states beyond the WF positivity ball is described.

\paragraph{Acknowledgments.}
The work is supported in part by the Bulgaria-JINR Program of Collaboration. 
One of the authors (A.K.) acknowledges the financial support of the  Shota Rustaveli National Science Foundation of Georgia, Grant FR-24-075. 
The work of A.T. was supported in part by the Higher Education and Science Committee of The Ministry of Education, Science, Culture and Sports of the Republic of Armenia (research project No 23/2IRF-1C003).



\begin{thebibliography}{99}
\bibitem{Deneris2025} A.E. Deneris  et al. (2025) Analyzing the free states of one quantum resource theory as resource states of another. https://arxiv.org/abs/2507.11793. 
\bibitem{Stratonovich1957}R.L. Stratonovich (1957)  On distributions in representation space. \textit{Soviet Physics JETP} \textbf{4}(6) 891-8.
\bibitem{BrifMann1999}C. Brif and A. Mann (1999) Phase-space formulation of quantum mechanics and quantum-state reconstruction for physical systems with Lie-group symmetries. \textit{Phys.Rev.} \textbf{59} 971-87. 
\bibitem{KhA2018}A. Khvedelidze and V. Abgaryan (2021) On families of Wigner 
functions for N-Level quantum systems. \textit{Symmetry} \textbf{13}(6) 1013.
\bibitem{AKH2023}A. Khvedelidze (2023) Generalizing Stratonovich–Weyl axioms for composite systems. \textit{Physics of Particles and Nuclei} \textbf{54}(6) 1025-8.  
\bibitem{AKHR2025}A. Khvedelidze and I. Rogozhin (2025) On the dichotomy between quatrits and pairs of qubits in the Stratonovich-Weyl correspondence.  \textit{Physics of Particles and Nuclei} \textbf{56}(4) 1025-9. 
\bibitem{BZ}I. Bengtsson and K. Zyczkowski (2017) Geometry of Quantum States: An Introduction to Quantum Entanglement (2nd ed.). Cambridge University Press. 
\bibitem{KhT2023}A. Khvedelidze and A. Torosyan (2023) On the hierarchy of classicality and symmetry of quantum states. \textit{Zap. Nauchn. Semin.} \textbf{528} 238-60. 
\end{thebibliography}
\end{document}